\documentstyle[12pt]{article}

\newcommand{\nc}{\newcommand}
\nc{\beq}{\begin{equation}}
\nc{\eeq}{\end{equation}}
\nc{\beqa}{\begin{eqnarray}}
\nc{\eeqa}{\end{eqnarray}}

\input epsf
\newwrite\ffile\global\newcount\figno \global\figno=1

\def\writedef#1{}
\def\figin{\epsfcheck\figin}\def\figins{\epsfcheck\figins}
\def\epsfcheck{\ifx\epsfbox\UnDeFiNeD
\message{(NO epsf.tex, FIGURES WILL BE IGNORED)}
\gdef\figin##1{\vskip2in}\gdef\figins##1{\hskip.5in}% blank space instead
\else\message{(FIGURES WILL BE INCLUDED)}%
\gdef\figin##1{##1}\gdef\figins##1{##1}\fi}
\def\figinsert{}
\def\ifig#1#2#3{\xdef#1{fig.~\the\figno}
\writedef{#1\leftbracket fig.\noexpand~\the\figno}%
\figinsert\figin{\centerline{#3}}\medskip\centerline{\vbox{\baselineskip12pt
\advance\hsize by -1truein\center\footnotesize{  Fig.~\the\figno.} #2}}
\bigskip\endinsert\global\advance\figno by1}
\def\endinsert{}
\begin{document}

\title{\Large{\bf 
The Field Theory of Non-Supersymmetric Brane
    Configurations }}

\author{  
%Yale Bulldogs and BU Terriers
%\\
Nick Evans\thanks{nevans@physics.bu.edu}
\\ 
{\small Department of Physics, Boston University, Boston, MA 02215} \\ \\
and\\ \\
Myckola Schwetz\thanks{myckola@baobab.rutgers.edu}
\\ {\small Department of Physics, Yale University, New Haven, CT 06520-8120}
\\
{\small and} \\ 
{\small Department of Physics and Astronomy,
Rutgers University, Piscataway, NJ 08855-0849}
}

\date{}

\maketitle

\begin{picture}(0,0)(0,0)
%\put(350,370){draft version}
\put(350,395){BUHEP-97-25}
\put(350,365){YCTP-P16-97}
\put(350,380){RU-97-70}
%\put(350,385){hep-th/97?????}
\end{picture}
\vspace{-24pt}

\begin{abstract}
We identify  the 4D field theories living on the
world volume of D4 branes in non-supersymmetric type IIA string theory
constructions. They are softly broken  N=2 SQCD with the breakings
introduced through vevs of the auxilliary fields in the
spurion coupling field. Exact solutions of these theories for 
perturbing soft breakings exist in the literature. We calculate the
ratios of string tensions in softly broken N=2 SU(N) gauge theory
testing the recently proposed M-theory prediction. The semi-classical
result of M-theory is renormalized in the non-supersymmetric models. 
\end{abstract}

\newpage
\section{Introduction}

The latest twist in the tale of the deciphering of supersymmetric gauge
theories has been provided by string theory. It has been realized that
4D gauge theories may be constructed as the effective theory on the
world volume of D-branes in string theory \cite{1} \cite{2}-\cite{15}
(an alternative realization of the field theory dualities in string
theory has been investigated in \cite{6}). 
The moduli spaces of
the SQCD theories have in this way been realized geometrically in the
brane language. Initially interest focused on realizing the
electro-magnetic dualities of N=1 SQCD geometrically in these
configurations \cite{3}-\cite{14}. 
The connection was then made between the brane
configurations and the IR solutions of N=2 SQCD \cite{4} 
\cite{7}-\cite{8};  in  particular, Witten \cite{7}
realized that by extending the decription of the type IIA configuration
to M-theory, and hence providing a description of the strong coupling
regime of the type IIA theory, the curve describing the M-theory
configuration was precisely that describing the IR physics of the gauge
theory. This techique has then been extended to provide derivations of
the superpotentials of confining N=1 SQCD theories \cite{12} \cite{13}. 
As proposed in \cite{14}
the success of these methods may be thought of as analogous to the
strong coupling expansion on the lattice. As one moves to M-theory from
the Type IIA theory the eleventh compactified dimension's radius is
increased from zero and its associated Kaluza Klien modes fall in mass
from infinity \cite{12}. 
Simultaneously the UV coupling increases and the M-theory
configuration provides a semi-classical solution of the strongly coupled
model. The solutions are thus for field theories with extra
states but which we hope nevertheless lie in the same universality class
as the SQCD theories we wish to study. The non-renormalization theorems
of supersymmetric theories preserve the semi-classical results into the
quantum theory. 

The latest success of the string theory approach has been to identify
the SQCD string \cite{12} and the confined bound states \cite{15}
in the (M-theory modified) SQCD theories with M-theory branes ending on
the configurations surface. The M-theory predictions for string tensions
are in agreement with the field theory results when N=2 SQCD is broken
to N=1 \cite{15}. The success in qualitatively identifying the theories bound
states in some sense supersedes the field theory successes.

The toughest challenge for these new methods is to move beyond SQCD to
non-super- symmetric theories. Some progress has been made within the
context of field theory in studying the
SQCD solutions in the presence of perturbing soft supersymmetry breaking
interactions \cite{16}-\cite{24} such as a gaugino mass. 
These breakings are introduced through the vevs of the
auxilliary components of spurion coupling fields \cite{16}, a technique that
allows the symmetries of the SQCD theories to be preserved into the
non-supersymmetric regime. Unfortunately the decoupling limit for the
super partners may not be taken, in N=1 SQCD because of unknown Kahler
terms that contribute directly to the potential of the softly broken
theories \cite{17}, 
and in N=2 SQCD because of the restrictions of the derivative
expansion of the SQCD solution enforcing an expansion in the breaking
terms over $\Lambda$ \cite{18} \cite{20}. 
Nevertheless these softly broken theories exhibit
different behaviour to the SQCD theories. The soft breakings break the
$Z_N$ symmetry of SQCD and the theories have a unique vacuum. 
They also display $\theta$-angle dependence \cite{17} \cite{19}, 
typically having phase
transitions at $\theta_{phys} =  ({\rm odd}) \pi$. 
%In one case \cite{21} where N=2 is broken directly to N=0 a transition to a
%vacuum that is pinned on the N=2 moduli space far away 
%from the position that N=1
%breakings pin the theory has been observed for controlled values of the
%soft breaking. 

Non-supersymmetric D-brane configurations have been proposed in Type IIA
string theory \cite{5} \cite{12} 
and Witten has provided a curve describing the M-theory
extension of that configuration. The first attempt at a quantitative
understanding of this set up has appeared recently in \cite{15}. 
The authors again
construct M-theory branes connecting the surface that may be
interpreted as QCD strings and semi-classically calculate  the string
tensions. They suggest that the ratios of string tensions found in the
N=1 SQCD theories might carry over to the N=0 theory.

In this paper we propose an identification of the field theory on this
N=0 brane configuration. We expect the breaking of 
N=2 supersymmetry in 10D string
theory by the string dynamics to appear as spontaneous
breaking in the low energy field theory description. Non-renormalizable
operators will be suppressed by the scale of the breaking (here given by
the brane tensions and hence of order the Planck mass). Thus any
breakings in the low energy theory 
will be precisely of the form of soft breakings that may be
introduced through the vevs of auxilliary spurion fields. The N=2 theory
has a single spurion coupling, the gauge coupling $\tau$ which is a
member of an N=2 spurion multiplet and hence there is a unique fashion
in which the soft breakings may enter the theory. 
We show that the induced breakings
correctly leave the field content described by the brane
configuration. 

The identification of the string theory configurations with softly broken
SQCD for which field theory solutions exist (for small breakings) allows
us to test the quantitative prediction of M-theory for the string
tensions in these models. We explicitly calculate the string tensions in
the SU(N) theory as one moves away from the supersymmetric point. The
semi-classical M-theory prediction is found to be only the zeroth order
contribution to the ratio of string tensions which have additional
contributions that can be expanded in powers of the soft breaking
parameter. We evaluate the ratios to first order in that breaking
parameter.

\section{Review of N=2 and N=1 Configurations}

We begin with a brief overview of the brane configurations describing
supersymmetric field theories (in this paper we restrict our attention
to models  without matter fields). The
constructions all consist of Type IIA  D4-branes suspended between NS5
branes which lie at various angles to each other. 

\subsection{N=2 SQCD}

The N=2
SU(N) gauge theory may be realized by the construction \cite{7}-\cite{8}
summarized as

\beq \label{N=2config}
\begin{tabular}{|c|c|c|c|c|c|c|c|c|}
\hline
 & $\#$ & $R^4$ & $x^4$ & $x^5$ &  $x^6$ & $x^7$ & $x^8$ & $x^9$ \\
\hline 
NS & 2 & $-$ & $-$ & $-$  &  $\bullet$ & $\bullet$ & $\bullet$ & $\bullet$ \\
\hline
D4 & $N_c$ & $-$  & $\bullet$ &  $\bullet$ &  $[-]$ & $\bullet$ & 
                              $\bullet$ & $\bullet$ \\
\hline
\end{tabular} 
\eeq

$R^4$ is the space $x^0-x^3$. A dash $-$ represents a
direction along a brane's world wolume while a dot $\bullet$ is
transverse. For the special case of the D4-branes' $x^6$ direction,
where the world volume is a finite interval corresponding to its
suspension between the two NS5 branes at different values of $x^6$, 
we use the symbol $[-]$. 

The configuration has N=2 supersymmetry in four dimensions. The 4D gauge
theory is realized on the $R^4$ world volume of the D4 branes.
Strings connecting the $N_c$ D4-branes give rise to gauge bosons on the
D4-branes' world volumes. When the D4-branes are coincident the gauge
group is an unbroken $SU(N_c)$ \cite{7}. The D4-branes can be separated in the
$x^4$ and $x^5$ directions in which case the elementary strings are no
longer of zero length and so correspond to massive gauge fields. In
general when the D4-branes are separated the gauge symmetry is
$U(1)^{N_c-1}$. 
Hence this motion is the geometrical representation of the higgs
mechanism associated with an adjoint scalar in the N=2 theory. 
The field theory exists on scales much greater 
than the $L_6$ distance between the NS5 branes with  the fourth
space like direction of the D4-branes generating  the coupling of the
gauge theory in the  effective 4D theory. 

Witten's IR solution of the model \cite{7} 
comes from extending the configuration
to M-theory by bringing the compactification radius, $R$,  of the eleventh
dimension $x^{10}$ up from zero. Whilst the type IIA string theory
description of the D4-branes ending on the NS5 branes is a strongly
coupled problem it has a smooth M-theory description since both the
D4-branes and NS5 branes are aspects of a single M-theory 5-brane but
which in places is wrapped around the compact $x^{10}$ direction. Using
the compact notation
\begin{equation}
v = x^4 + i x^5, \hspace{1cm} t = exp{(x^6+ix^{10})/R}
\end{equation}
The holomorphic curve $\Sigma$ describing the field theory configuration
above is
\begin{equation} \label{N=2curve}
t^2 + B(v)t + 1 = 0
\end{equation}
where $B(v)$ is a polynomial of order $N_c$ in $v$. The curve is
precisely that of the form of the IR solution of the N=2 field theory.

Recall that the N=2 field theory \cite{9}-\cite{11} has a quantum moduli space
corresponding to different choices of the adjoint vev. The gauge
symmetry is generically broken down from $SU(N_c)$ to
$U(1)^{N_c-1}$ and the neutral components of the adjoint matter field
remain massless. 
There are $N_c$ points on the moduli space with $N_c-1$ 
massless monopoles or dyons.  At each point, in the
basis where all the local low energy $\theta$ angles are set to zero, there are
$N_c-1$ massless monopoles each charged only under a single $U(1)$
factor. The electric variables are strongly coupled about these points
but the electro-magnetic
 duality of the theory provides a weakly coupled dual description
of the physics about each point. The electric and magnetic variables for
the $N_c-1$ massless components of the adjoint matter field, 
$a_i$ and $a_{Di}$ are
given by period integrals of the curve (\ref{N=2curve}). These periods can be
translated to the more useful form of a prepotential appropriate to the
local patch of the moduli space \cite{11}. For example the theory close to a
singular point is described by the Lagrangian
\begin{eqnarray} 
{\cal L} & = &\sum_i^{N_c-1} \left( {1 \over 4 \pi} Im 
\left[ \int d^4\theta {\partial {\cal
    F}_{\rm eff} \over \partial A_{Di}} \overline{A}_{Di} + {1 \over 2} 
\int d^2\theta
{\partial^2 {\cal F}_{\rm eff} \over \partial A_{Di}^2} W^i 
W^i \right] \right. \nonumber \\
&& \left. + \int d^4\theta (M^\dagger_i e^{2V_{Di}}M_i +
\tilde{M}_i^\dagger e^{-2V_{Di}} \tilde{M}_i ) + 2 \sqrt{2} Re \int A_{Di} M_i
\tilde{M}_i \right)
\end{eqnarray}
where ${\cal F}_{\rm eff}$ is the prepotential \cite{11}
\begin{equation} \label{prepotential}
{\cal F}_{\rm eff} = -i {N_c^2 \Lambda^2 \over 2 \pi} 
-{2N_c \Lambda \over \pi} \sum_{k=1}^{N_c-1} s_k A_{Dk} - {i \over 4\pi}
\sum_{k=1}^{N_c-1} a^2_{Dk} \ln {a_{Dk} \over \Lambda_k} + {\cal O}(a_D^3)
\end{equation}
with 
\beq
s_k = \sin {k \pi \over N_c}, \hspace{1cm} \ln {\Lambda_k \over \Lambda}
= {3 \over 2} + \ln s_k
\eeq 
and
\beq
\Lambda = \Lambda_{UV} e^{i 2 \pi \tau_0 / b_0}, \hspace{1cm} \tau_0 =
{\theta_0 \over 2 \pi} + i {4 \pi \over g_0^2}
\eeq
where $b_0 = 2 N_c$ is the one loop $\beta$ function coefficient and $g_0$ the UV
coupling at scale $\Lambda_{UV}$.

We note that at higher order in $a_{Dk}$ the $U(1)$ factors begin to mix
in the IR effective theory (in all calculations below we will work at
lower order and neglect this complication).

\subsection{N=1 SQCD}

N=1 SQCD is realized very similarly to the N=2 configuration but by
simply rotating one of the NS5 branes into the $x^8-x^9$
directions. There is in fact a continuous set of N=1 configurations \cite{7.5}
associated in the brane language with rotating the ``$v$-plane'' of one NS5
brane in the N=2 configuration into the $x^8-x^9$ directions. In the
field theory this is interpreted as smoothly turning on a mass for the
adjoint matter field of the N=2 model. The configuration is now

\beq \label{N=1config}
\begin{tabular}{|c|c|c|c|c|c|c|c|c|}
\hline
 & $\#$ & $R^4$ & $x^4$ & $x^5$ &  $x^6$ & $x^7$ & $x^8$ & $x^9$ \\
\hline 
NS & 1 & $-$ & $-$ & $-$  &  $\bullet$ & $\bullet$ & $\bullet$ & $\bullet$ \\
\hline
NS & 1 & $-$  & $\bullet$ & $\bullet$  &  $\bullet$ & $\bullet$ & $-$ & $-$ \\
\hline
D4 & $N_c$ & $-$  & $\bullet$ &  $\bullet$ &  $[-]$ & $\bullet$ & 
                              $\bullet$ & $\bullet$ \\
\hline
\end{tabular} 
\eeq

The IR superpotential is again obtained by moving to the M-theory
description \cite{12} in which the curve is described by (with $w=x^8 +i x^9$)
\begin{equation}\label{N=1curve}
t = v^{N_c}, \hspace{1cm} vw = \xi
\end{equation}

For large $v$, $w \simeq 0$ and $\Sigma$ has a $Z_{N_c}$ symmetry but
in the interior of the curve the symmetry is broken by the second
equation in (\ref{N=1curve}).
The IR superpotential, $W = N_c \Lambda^3 \simeq 4 \pi i R N_c \xi$ 
may be obtained from $\Sigma$ by performing a
volume integral in $v,w, ln\,t$ space over an appropriate volume with
$\Sigma$ as one surface (see \cite{12}).

\section{N=0 Branes And Field Theory Identification}

Our interest here is in the possibility that these brane configurations
may be able to shed light on non-supersymmetric configurations. We will 
again only consider the matter free case
in the hope of identifying the resulting field theories. As
pointed out in \cite{12} an arbitrary rotation of the $v$-plane
of one NS5 brane of the N=2 configuration in the $x^4, x^5, x^7, x^8,
x^9$ volume
leads to an N=0 configuration. There are 6 degrees of freedom
($SO(5)/SO(2) \times SO(3))$ associated with these rotations. 
Witten has
provided a discription of the M-theory configuration corresponding to
models with arbitrary rotations of the $v$-plane in the $x^4, x^5, x^7,
x^8$ volume (we write these four coordinates as the vector $\vec{A}$). 
It is a minimal area embedding solution and takes the form
\begin{eqnarray}\label{0brane}
\vec{A} &  = & Re(\vec{p} \lambda + \vec{q} 
\lambda^{-1}) \nonumber \\
x^6 & = & -cRN_c Re \ln \lambda \\
x^{10} & = & - N_c Im \ln  \lambda \nonumber
\end{eqnarray}
where $\vec{p}$ and $\vec{q}$ are complex 
four vectors and with $c$ are chosen
to satisfy the Virasoro constraint
\beq
\vec{p}^2 = \vec{q}^2 =0,\hspace{1cm} 
-\vec{p}.\vec{q} + {R^2 N_c^2 \over 2}(1-c^2) =0 
\eeq

The supersymmetric configurations of (\ref{N=2config}) and
(\ref{N=1config}) are realized by the choices: $\vec{p}=(1,i,0,0),
\vec{q}=(1,i,0,0), c=\pm 1$ (N=2), and  $\vec{p}=(1,i,0,0),
\vec{q}=(0,0,1,i), c=\pm 1$ (N=1). In addition the N=0 configuration
of \cite{5} may be reproduced ($\vec{p}=(1,i,0,0),
\vec{q}=(1,0,i,0), c=\pm (1-2/N_c^2R^2)^{1/2}$)

\beq \label{N=0config}
\begin{tabular}{|c|c|c|c|c|c|c|c|c|}
\hline
 & $\#$ & $R^4$ & $x^4$ & $x^5$ &  $x^6$ & $x^7$ & $x^8$ & $x^9$ \\
\hline 
NS & 1 & $-$ & $-$ & $-$  &  $\bullet$ & $\bullet$ & $\bullet$ & $\bullet$ \\
\hline
NS & 1 & $-$  & $-$ & $\bullet$  &  $\bullet$ & $-$  &
$\bullet$   & $\bullet$ \\
\hline
D4 & $N_c$ & $-$  & $\bullet$ &  $\bullet$ &  $[-]$ & $\bullet$ & 
                              $\bullet$ & $\bullet$ \\
\hline
\end{tabular}
\eeq

This configuration describes a non-supersymmetric 
$SU(N_c)$ gauge theory with a real
adjoint scalar field corresponding to the freedom to separate the D4-branes
in the $x^4$ direction. 

Can we identify the supersymmetry breaking terms
introduced into the N=2 theory on the brane?
The string theory at tree level is a supersymmetric theory. Thus when we
find a low energy description in which supersymmetry is broken it can
only be because the background field configurations (the branes) have induced
supersymmetry breaking through the vevs of fields in the fundamental
theory. Supersymmetry breaking is generated by vevs of the auxilliary
fields of chiral superfields. 
The low energy supersymmetry breaking parameters are
therefore given by these field vevs in a low energy description in which
oscillations about those vevs are ignored. Such fields are called
spurion couplings in the field theory. A detailed discussion of this
mechanism in string theory where the supersymmetry breaking is the
result of a choice of compactification is in \cite{20}. 
In the brane picture these fields also
have explicit realizations. For example the gauge coupling is given by
the $x^6$ distance between the two NS5 branes which can equally be
thought of as the vev of a scalar field living on the branes
(describing the brane positions). The
oscillations of the scalar are neglected since the branes are so heavy
that any oscillation in the $x^6$ direction is negligible in the low
energy field theory.

If we do not neglect those oscillations then the coupling will appear in
the field theory as a field. This scalar as part of a superfield will
have superpartners. For the N=2 configuration these partners will fill
out an N=2 multiplet. In fact as discussed in \cite{20} there is a
unique way in which the coupling may be included in the field theory as
a spurion field. It must appear as a vector multiplet corresponding to a
U(1) symmetry \cite{20}. In fact in complete generality there can be two
spurion fields that enter the 4D field theory prepotential as
\beq \label{prepot}
{\cal F}_{cl} = {N_c \over 2  \pi} (S_1 + i S_2) A^2
\eeq
One might be tempted to write $S_1$
and $S_2$ as sums or products of superfields but the spurion 
symmetries are such as to constrain those fields to only ever occur in
the combinations $S_1$ and $S_2$. One might also include terms of higher
dimension in $A$ (the scalar spurion vev which is the coefficient 
must then be zero in the N=2 limit so
these terms vanish) but these terms will generate higher dimension terms in
the Lagrangian which are irrelevant for the low energy physics.
The Lagrangian is given by
\begin{eqnarray} \nonumber
{\cal L} &~=~& \frac{1}{4 \pi} Im \left[ \int d^4 \theta  \left(
\frac{\partial {\cal F}_{cl}}{\partial A} \bar{A} ~+~
\frac{\partial {\cal F}_{cl}}{\partial S_1} \bar{S_1}  ~+~
\frac{\partial {\cal F}_{cl}}{\partial S_2} \bar{S_2} \right)\right.\\
&&\left. + 
\int d^2 \theta \left( 
{1 \over 2} \frac{ \partial^2 {\cal F}_{cl}} {\partial A^2} W
W
+
 \frac{ \partial^2 {\cal F}_{cl}} {\partial A \partial S_1} W W'
+
{1 \over 2} \frac{ \partial^2 {\cal F}_{cl}} {\partial S_1^2} W' W' \right.
\right.\\
&&+ \left. \left.
 \frac{ \partial^2 {\cal F}_{cl}} {\partial A \partial S_2} W W^{''}
+
{1 \over 2} \frac{ \partial^2 {\cal F}_{cl}} {\partial S_2^2} W^{''} W^{''} +
 \frac{ \partial^2 {\cal F}_{cl}} {\partial S_1\partial S_2 } W^{'} W^{''}
 \right)
\right]~.\nonumber
\end{eqnarray}
Freezing  the scalar components of the spurion
multiplets' matter fields generates the coupling
of the pure glue model with $s_1 + i s_2= \pi \tau_{0}/N_c$ (this normalization
is convenient since $\Lambda \sim exp(i(s_1 + is_2))$).

Since this is the unique way in which a spurion may enter the N=2 SQCD
theory there is a unique way in which soft breakings may be
included. 
That is by also freezing the complex $F$-components   
of the spurion matter fields,
$F_1, F_2$, or  the real $D$-components   of the spurion vector fields,
$D_1, D_2$, and generate soft breaking masses 
\begin{eqnarray} \label{softUV}
&& \nonumber
-{N_c \over 8  \pi^2} Im \left( (F_1^*+iF_2^*) \psi_A^\alpha \psi_A^\alpha 
+ (F_1 + i F_2)
\lambda^\alpha \lambda^\alpha + i \sqrt{2}
(D_1 + i D_2) \psi_A^\alpha\lambda^\alpha \right) \\
&&
  - {N_c \over 4 \pi^2 Im (s_1+is_2)} \left((|F_1|^2 + D_1^2/2) Im(
a^\alpha)^2
+  (|F_2|^2 + D_2^2/2) Re
(a^\alpha)^2 \right. \\
&& \left. \right. \hspace{4cm} + \left. 
 (F_1 F_2^* + F_1^*F_2 + D_1D_2) Im(a^\alpha) Re(a^\alpha)
\right)\nonumber
\end{eqnarray}

As there is a unique possibility for how these spurions occur in the
field theory the string theory brane configurations have no choice but
to break supersymmetry in this fashion. A number of checks can be
performed comparing the field theory and brane configuration that seem
to bare out this statement. Firstly there are six independent degrees of
freedom that break supersymmetry in each case. In the field theory $F_1,
F_2, D_1, D_2$. In the brane configuration there are six independent
rotations of one of the NS5 branes that break N=2 supersymmetry to N=0 
as can be seen from (\ref{N=2config}) and
(\ref{N=0config}); the $x^4$ dimension into each of
$x^7,x^8,x^9$ or the $x^5$ dimension into  $x^7,x^8,x^9$. These
rotations each leave a real scalar degree of freedom massless
corresponding to the ability to shift D4 branes in the unrotated one of
those two coordinates. Encouragingly switching on any one of the field
theory parameters also leaves a massless scalar as can be observed in
(\ref{softUV}). The identification
between the brane picture and the field theory is therefore that each of
these six rotations corresponds to switching on one of the soft breaking
parameters. 

In the brane picture rotating both of the $x^4,x^5$ directions together
into the $x^6, x^7$ plane can be performed in a N=1 supersymmetry preserving
way \cite{7.5}. If our identification is correct then this
should correspond to switching on two of the soft breaking parameters
with equal magnitudes. We indeed find in the field theory that this
leaves an N=1 supersymmetric spectrum; for example setting
$Re(F_2)=Im(F_1)  = M$ and all other components zero leaves the gaugino
in the field theory massless. The massive fermion and the two components
of the scalar have a mass consistent with adding to the N=2 theory a term 
\beq 
{N_c \over 4  \pi^2}  Im \int  d^2 \theta  M  A^2
\eeq

These non-trivial consistency checks strongly suggest that the field
theory identification is correct.

\section{The QCD Strings}

Witten \cite{12} has identified the SQCD string in the M-theory broadened
configurations corresponding to N=1 SQCD 
with M-theory two-branes ending on $\Sigma$ so they
appear as strings in $R^4$. They will take the form of a one-brane on
$R^4$ ``crossed'' with a one-brane outside $R^4$ with each end on
$\Sigma$. The string outside $R^4$ can be
described by
\begin{eqnarray} \label{string}
v & = & \sqrt{\xi} t_0^{1/N_c} {\rm exp} ( 2 k \pi i \sigma/ N_c) \nonumber \\
w & = & \xi v^{-1} 
\end{eqnarray}
where $t_0^{1/N_c}$ is a particular $N_c$th root of $t_0$. $\sigma$ is a
coordinate along the string and runs from 0 to 1. For
each choice of $k = 1..N_c-1$ the string wraps $k$ times in $x^{10}$
(a string that wraps $N_c$ times has been shown by Witten to vanish
corresponding to the annihilation of $N_c$ mesons into two baryons). The
SQCD string tensions are therefore proportional to the mimimal possible 
length of the
string in (\ref{string})
\beq
\sqrt{ |\Delta v|^2 + |\Delta w|^2} = 2 \sqrt{\xi} \sqrt{t_0^{2/N_c} +
  t_0^{-2/N_c}} \sin(\pi k /N_c)
\eeq
The minimum of this length is when $t_0=1$.
These theories therefore have $N_c-1$ different SQCD strings with ratios
of string tensions \cite{15}
\beq \label{stratio}
{T_k \over T_l} = {\sin {\pi k \over N_c} \over \sin{\pi l \over N_c}}
\eeq
In \cite{15} it has been shown that this result holds for the complete
set of N=1 brane configurations obtained by smoothly rotating the $v$-plane of
one NS5 brane into the N=2 configuration. 
This result is in agreement with the calculated string tensions in
models where N=2 $SU(N_c)$ SQCD is broken to N=1 with adjoint matter
mass \cite{10}.

A similar M-theory calculation can be carried out for strings in the N=0
configurations \cite{15}. 
The QCD strings are again two-branes in the M-theory. By the symmetry of
the brane set up the minimal length strings outside $R^4$ connecting
$\Sigma$ and wrapping $k$ times around $x^{10}$ will lie at $x^6=0$ and
hence the end points will lie at the points parameterized in
(\ref{0brane}) by $\lambda = e^{i\sigma_1}, e^{i\sigma_2}$, with
$\sigma_1 - \sigma_2 = 2\pi k/N_c$. The ratios of string lengths is again
given by (\ref{stratio}).
The authors of \cite{15} proposed that
this semi-classical result may hold in the N=0 quantum theory. We test
this quantitative prediction in the next section.

\section{N=0 Field Theory Calculation of String Tensions}

We have identified the field theories corresponding to the low energy
description of the N=0 brane configurations as N=2 SQCD softly broken by
the N=2 $\tau$ spurion vevs. These models have been studied and solved
\cite{20}-\cite{22}
for small perturbing soft breakings (corresponding to small angle
rotations relative to the starting N=2 configuration in the brane
picture). We will make use of these solved models to test the M-theory
predictions for the ratios of string tensions described above.

The quantum theory of the softly broken models is described by a
prepotential that is holomorphic in $A_{i}$ and $S_j$ (below we denote $S_1
= A_{D1}$, $S_2=A_{D2}$ and the $N_c-1$ U(1) sectors by the index $i$). 
The occurence of the spurions  is known from the N=2
solutions since their lowest component vevs are simply $\tau_0$. That is
the prepotential remains that of (\ref{prepotential}).
Note that $S_i$ are
globally defined variables  
while $A_i$ and $A_{Di}$
are only locally defined on the N=2 moduli space. The N=0 theory with
$N_c=2$ has been completely studied \cite{20} and the global minimum
lies close to the singular point on the N=2 moduli space with a
massless monopole. For larger $N_c$ the theory is strongly believed 
\cite{21} to
behave similarly pinning the potential  near the singular point with
$N_c-1$ massless monopoles. We shall study the theory in the local
coordinates appropriate to that patch of parameter space. The theory is
described by
\begin{eqnarray}
{\cal L}& ~=~&\sum_{i,j} \left( 
 \frac{1}{4 \pi} Im  \int d^4 \theta  
\frac{\partial {\cal F}_{\rm eff} }{\partial A_{Di}} \bar{A}_{Di} 
+ \frac{1}{8 \pi} Im
\int d^2 \theta \left( \frac{ \partial^2 {\cal F}_{\rm eff}} 
{\partial A_{Di} \partial A_{Dj}} W_i W_j 
\right)  \right. \nonumber \\
&& \left.  + \int d^4\theta M^\dagger_i e^{2V_{Di}}M_i +
\tilde{M}_i^\dagger e^{-2V_{Di}} \tilde{M}_i  + 2 \sqrt{2} Re \int A_{Di} M_i
\tilde{M}_i  \right)
\end{eqnarray}
with ${\cal F}_{\rm eff}$ given by (\ref{prepotential}). 
Performing the superspace
integrals \cite{21} we obtain the potential 
\begin{eqnarray}
V & = & \sum_i  \left[  {1 \over 2 b_{ii}} 
(|m_i|^2 + |\tilde{m}_i|^2)^2 +
2|a_{Di}|^2(|m_i|^2 + |\tilde{m}_i|^2)\right. \nonumber \\
&& + {1 \over b_{ii}} \left( \sqrt{2} (b_{1i}F_1^*+b_{2i}F_2^*) m_i
  \tilde{m}_i + \sqrt{2} (b_{1i}F_1 + b_{2i}F_2) 
\overline{m}_i \overline{\tilde{m}}_i  \right)\nonumber \\
&&
+ {1 \over b_{ii}}
(b_{1i}D_1 + b_{2i}D_2) 
(m_i|^2
- |\tilde{m}_i|^2) \nonumber \\
&&
+ \left. {1\over b_{ii}} \left(b_{1i}^2(|F_1|^2+{1\over 2} D_1^2)
 + b_{2i}^2 (|F_2|^2 + {1\over 2} D_2^2)+ b_{1i}b_{2i} (F_1F_2^* + F_1^*
   F_2 + D_1D_2) \right)    \right] \nonumber \\
 && -  b_{11} (|F_1|^2+{1 \over 2} D_1^2)  
- b_{22} (|F_2|^2 + {1\over 2}D_2^2)  - b_{12}(F_1F_2^* + F_1^* F_2 + D_1D_2)
\end{eqnarray}
where
\beq
b_{ij} = {1 \over 4 \pi}Im {\partial^2 {\cal F}_{\rm eff} \over \partial
  a_i \partial a_j }
\eeq

For simplicity we set the spurion components $D_1=D_2 = 0$. Minimizing
with respect to $m_i$ and $\tilde{m}_i$ leads to the result 
(see \cite{21} for details)
\begin{eqnarray} 
\label{pot}
 V &~=~& - b_{11}|F_1|^2 - b_{22} |F_2|^2 - b_{12} (F_1^*F_2 + F_1
F_2^*) \nonumber \\
&&  +  \sum_i  \Bigg[ 
 {b_{1i}^2 \over b_{ii}}  |F_1|^2
~+~ {b_{2i}^2 \over b_{ii}}|F_2|^2
~+~ {b_{1i}b_{2i} \over b_{ii}} (F_1^*F_2 + F_1F_2^*)  ~-~
2\, {\rho_i^4 \over b_{ii}} \,\Bigg]
\end{eqnarray}
and
\beq \rho^2_i = |m_i|^2 = | \tilde{m}_i|^2 
= - b_{ii} |a_{Di}|^2 - {|b_{1i}F_1 + b_{2i} F_2|
  \epsilon_\pm
\over \sqrt{2}}
\eeq
where $\epsilon_\pm$ is dynamically determined between its two possible
values $\pm 1$.

We may now find the minimum of the potential using the form of the
prepotential in (\ref{prepotential}).
For  most choices of soft breakings and $\Lambda$ it is sufficient to work
to order $|a_{Di}|^2$ to find the minimum. For special choices of
parameters, for example taking $\Lambda$ real ($\theta=0$) and only $F_2
\neq 0$, the $|a_{Di}|^2$ terms in the potential can be made to vanish
and one must go to order $|a_{Di}|^4$. It is sufficient for
displaying the behaviour of the models to work at order  $|a_{Di}|^2$
and we find
\beq 
\langle a_{Di} \rangle = {i \Lambda  \over 4 \sqrt{2}}{ \left( |F_1|^2
  - |F_2|^2 + i (F_1^*F_2 + F_1F_2^*)\right) \over ( |Re(\Lambda) F_1 -
  Im(\Lambda) F_2|)}
\eeq

Let us now consider some special cases of this result. Firstly when
$F_2=iF_1 = iM$ the tree level theory was N=1 supersymmetric. The IR solution
is just the familiar result of \cite{10}, $\langle a_{Di} \rangle = 0$
(which can be shown to all orders from (\ref{pot})) and hence 
\beq
\rho_i = {\sqrt{2} N_c \over 4 \pi^2} s_i \Lambda^* M
\eeq
The string tensions are given by \cite{25}
\beq 
T_k = 2 \pi | \langle m_k \tilde{m}_k \rangle | = 2 \pi |\rho_k|
\eeq
and are proportional to $\sin(k \pi/N_c)$ in agreement with the
M-theory calculation (\ref{stratio}).

We may also calculate the string tensions in the non-supersymmetric
theories. As an example we take $|F_1| = f_1$, $F_2 =0$ and $\theta
=0$. These theories
correspond to those in the string theory where the $v$-plane of one NS5
brane is rotated between the configurations of (\ref{N=2config}) and
(\ref{N=0config}). We can of course only keep control of the solution
for small soft breakings corresponding to small angles of rotation away
from the N=2 configuration. The minimum of the potential corresponds to
\beq 
\langle a_{Di} \rangle = {i f_1 \over 4 \sqrt{2}}, \hspace{1cm}
\rho_i^2 = {\sqrt{2} N_c \Lambda f_1 s_i \over 4 \pi^2} + {f_1^2 \over
  64 \pi^2} + {f_1^2 \over
  256 \pi^2}  \ln(f_1/4\sqrt{2}s_i \Lambda)
\eeq
The resulting ratios of string tensions is
\beq
\label{ratio}
{T_k \over T_l}~=~{s_k \over s_l}~+~ {\sqrt{2} \over 32 N \Lambda}
{s_l - s_k \over s^2_l}  \left[ 
 1 ~+~ {1\over 4}\, \ln{{f_1 \over 4 \sqrt{2}\Lambda}}   ~+~  
{1 \over 4}\, {s_k \ln {s_l } - 
s_l \ln{s_k } \over s_l - s_k} \right] f_1
~ +~ {\cal O} (f_1^2)~.
\eeq
The string tension prediction approaches the 
semi-classical M-theory prediction as $f_1 \rightarrow 0$ but is
renormalized in the softly broken theory. This is perhaps not so great a
surprise. As the authors of \cite{15} point out, the semi-classical results
are correct in the supersymmetric theories because supersymmetry
endows these quantities with a non-renormalization property. There is no
such property in the softly broken theories and a priori one should not
expect the semi-classical calculation to hold in the non-supersymmetric 
theory. In some sense though we view (\ref{ratio}) as a success of the
semi-classical M-theory calculation since the zeroth order term is the
M-theory result.

\section{Conclusions}

We have argued that N=0 D-brane configurations recently proposed
in the literature correspond to softly broken N=2 SQCD where the soft
breakings result from expectation values of the N=2 spurion coupling
$\tau$. These theories are solvable for small soft breakings
(corresponding to small rotations away from the N=2 configuration in
the brane language). A semi-classical calculation of the string tensions
in the theory can be made from the M-theory and agrees with the
supersymmetric field theory solutions. The equivalent calculation for
N=0 theories has been performed in \cite{15} and proposed as the correct
description of the quantum theory. {\it 
We stress that any such predictions are
testable in quantitatively solved non-supersymmetric field theories}. 
We have explicitly
calculated the ratio of string tensions in the N=0 field theories
 showing that at
leading order in the soft breaking the N=0 models renormalize the
M-theory result. Unfortunately this result seems to put pay to any hope
of obtaining quantitative results about QCD from the M-theory
construction.  

Finally we comment on possible lattice tests of
(\ref{ratio}) \cite{24} \cite{26}. Unfortunatly these will be difficult. Lattice
regularization explicitly breaks supersymmetry and, while one might hope
to recover supersymmetry in the continuum limit, in practice one
recovers the model with all possible breaking terms allowed by the gauge
symmetry. For pure glue N=1 SQCD the only possible such term is a
gaugino mass which is a soft breaking of supersymmetry \cite{lat}. In the N=2
theory though there is no such promise and we expect explicit hard
breakings to be present. To obtain the softly broken theories discussed
above would require the tuning of many bare lattice operators. Tuning to
the supersymmetric point would be equally troublesome but were it
identified one might hope to see the string tensions tending to the
M-theory ratio (our zeroth order result) as that point is
approached.\vspace{1cm}

\noindent {\large \bf Acknowledgements}: The authors would like to thank
R. Sundrum and S. Hsu for insightful comments on the manuscript. 
This work was in part funded by
the DOE under contract numbers DE-AC02-ERU3075  and DE-FG02-92ER40704
and by the NSF under grant number NSF-PHY-94-23002.

\newpage
% ------------------------------------------------------------
% Referencing
% ------------------------------------------------------------

% Journals
\newcommand{\jref}[4]{{\it #1} {\bf #2}, #3 (#4)}
\newcommand{\NPB}[3]{\jref{Nucl.\ Phys.}{B#1}{#2}{#3}}
\newcommand{\PLB}[3]{\jref{Phys.\ Lett.}{#1B}{#2}{#3}}
\newcommand{\PR}[3]{\jref{Phys.\ Rep.}{#1}{#2}{#3}}
\newcommand{\PRD}[3]{\jref{Phys.\ Rev.}{D#1}{#2}{#3}}
\newcommand{\PRL}[3]{\jref{Phys.\ Rev.\ Lett.}{#1}{#2}{#3}}
%---------------------------------------------------

\end{document}